\documentstyle[12pt,aasms4,psfig]{article}

\slugcomment{Submitted to the Astrophysical Journal Supplement Series}

\begin{document}

\title{The Edinburgh/Durham Southern Galaxy Catalogue -- IX.\\
    The Galaxy Catalogue}

\author{Robert C. Nichol\altaffilmark{1,3,4}} 
\affil{Department of Physics, Carnegie
Mellon University, 5000 Forbes Ave., Pittsburgh, PA--15213, USA}

\author{Christopher A. Collins\altaffilmark{1,2}} 
\affil{Astrophysics Research
Institute, Liverpool John Moores University, Twelve Quays House,
Egerton Wharf, Birkenhead L41 1DL, UK}

\author{Stuart L. Lumsden\altaffilmark{1,5}} 
\affil{Department of Physics and Astronomy,
University of Leeds, Leeds LS2 9JT, UK}

\altaffiltext{1}{Royal Observatory Edinburgh, Blackford Hill,
Edinburgh, EH9 3HJ, UK} 
\altaffiltext{2}{Department of Physics, Science Laboratories, South
Road, Durham DH1 3LE, UK} 
\altaffiltext{3}{Department of Physics \& Astronomy, Northwestern
University, Dearborn Observatory, 2131 N. Sheridan Road, Evanston,
IL-60208, USA}
\altaffiltext{4}{Department of Astronomy \& Astrophysics, University
of Chicago, 5640 S. Ellis Avenue, Chicago, IL-60637, USA}
\altaffiltext{5}{Anglo-Australian Observatory, P.O. Box 296, Epping,
NSW 1710, Australia}

\begin{abstract}

We announce here the public availability of the Edinburgh/Durham
Southern Galaxy Catalogue (EDSGC, http://www.edsgc.org).  This
objective galaxy catalogue was constructed using the COSMOS
micro-densitometer at the Royal Observatory Edinburgh and constitutes
one of the largest digitized galaxy surveys currently in
existence. The EDSGC contains a total of 1,495,877 galaxies (each with
27 image parameters) covering an contiguous area of $1182{\rm deg^2}$
centered on the South Galactic Pole. The data consists of photographic
$b_j$ magnitudes calibrated via CCD sequences which provide a
plate--to--plate accuracy of $\Delta b_j\simeq0.1$. Extensive external
checks have demonstrated that the global EDSGC photometry is free of
large--scale systematic gradients and is therefore, ideal for studying
the distribution of galaxies on large angular scales. Independent
spectroscopy of EDSGC galaxies has shown that the accuracy of the
star--galaxy separation is consistent with earlier visual checks and
that only 12\% of EDSGC galaxies are potentially mis--classified. This
paper is intended to provide a summary of the essential details of the
catalogue's design and construction as well as provide a brief summary
of the main scientific achievements using the EDSGC over the last ten
years.

\end{abstract}

\keywords{surveys -- galaxies: general -- galaxies: photometry -- cosmology: observations}

\section{Introduction}

Surveys of galaxies are invaluable tools for cosmologists.  To date,
the most comprehensive catalogues of galaxies have resulted from the
optical photographic surveys carried out with the Schmidt telescopes
at the Palomar and Anglo--Australian Observatories. The earliest of
these surveys were visually compiled counts of galaxies, the largest
being the Lick catalogue of galaxies which presents counts in cells of
$10\times10$ arcminutes for some 800,000 galaxies covering two thirds of
the entire sky. The complementary Abell cluster catalogue (Abell,
Corwin \& Olowin 1989) is an amalgamation of Abell's original northern
cluster catalogue (Abell 1958), carried out with the Palomar Sky
Survey 103a--E plates, with the southern extension using the IIIa-J
Southern Sky Survey plates taken by the UK Schmidt at Siding
Springs. The final all--sky Abell catalogue consists of 4073 rich
clusters of galaxies each having at least 30 members with a nominal
redshift of $z\leq0.2$ and is the first serious attempt to
systematically categorize the properties of galaxy clusters. The Abell
catalogue remains the most widely used cluster catalogue in the
astronomical literature.

The early 1990s saw the completion of the first large-scale digitized
galaxy surveys. These combined the best available photographic
material, in the form of UK Schmidt IIIa-J photographic survey plates,
with digitization technology in the form of fast micro-densitometers
{\it e.g.} the COSMOS machine in Edinburgh and the APM machine in
Cambridge. The digitization process overcomes the inherent
subjectivity of `eyeball' surveys and the impact of the resulting
galaxy catalogues has been extensive (see Section 3).

As we enter a new millennium, the number and size of galaxy surveys
will expand rapidly due to new photographic galaxy surveys ({\it e.g.}
DPOSS, Djorgovski et al. 1998; SuperCOSMOS, Phillipps et al. 1998) and
CCD surveys ({\it e.g.}  SDSS, York et al. 2000). These new surveys
will supersede the presently available galaxy catalogues because of
their increased sensitivity, areal coverage, color information and
more accurate photometry.  Therefore, in this paper, we announce the
public release of the Edinburgh/Durham Southern Galaxy Catalogue
(EDSGC) to assist cosmological research and is now particularly timely
given the completion of several redshift galaxy surveys for which the
EDSGC has been the input object catalogue (see Section \ref{sup}).

The structure of this paper is as follows: In Section \ref{cat}, we
discuss the construction of the EDSGC and describe the available
parameters and catalogue characteristics. In Section \ref{avail}, we
outline the availability of the EDSGC and in Section \ref{sup}, we
present a summary of the supplementary data available in the EDSGC
area. In Section \ref{check}, we present the extensive external checks
that have been performed on the catalogue, while in Section
\ref{science}, we review the science which has been carried out using
the EDSGC.

\section{The Edinburgh/Durham Southern Galaxy Catalogue}
\label{cat}

The Edinburgh/Durham Southern Galaxy Catalogue (EDSGC) is one of the
first fully automated objective galaxy catalogues to be constructed.
It covers an area of $\simeq1200{\rm deg^2}$ centered on the South
Galactic Pole (SGP) and contains extensive information on over one
million galaxies.  The majority of the initial work involved in its
construction has already been presented in several places {\it i.e.}
Collins, Heydon-Dumbleton \& MacGillivray 1989, Heydon-Dumbleton,
Collins \& MacGillivray 1989 and Heydon-Dumbleton 1989. In this
section we summarize the data and the methods used in the catalogue's
construction and presents some of the tests implemented on the
completed EDSGC.

\subsection{The Raw Data}
\label{raw}

The UK Schmidt Telescope (UKST) at Siding Springs in Australia was
commissioned in late 1973 to carry out a systematic photographic
survey of the southern hemisphere.  The UKST has a 1.8m diameter
mirror with a 1.2m aperture which provides a large field of view
making it ideal for the construction of such a large-area survey.  The
photographic plates used by UKST subtends an area of
$6.4^{\circ}\times6.4^{\circ}$ on the sky with a plate scale of 67.12
arcseconds per millimeter.  For all the surveys carried out by UKST,
the centers of the plates are separated by $5^{\circ}$ which provides
a substantial overlap between them and removes the need to use the
plate edges which are heavily vignetted (vignetting is negligible
within $2.7^{\circ}$ of the plate center).

The EDSGC is based on plates taken from the ESO/SERC Atlas.  This
atlas consists of glass copies of the SERC J survey which was the
first survey to be completed by UKST and covers the whole southern sky
below a declination of $-17^{\circ}$ (606 plates in total).  The
passband of the SERC J survey is defined by the response of the
emulsion (Kodak IIIA-J) combined with a Schott GG395 filter.  This
provides an almost uniform sensitivity in the wavelength range
$3500\AA$ to $5400\AA$ and is close to the standard Kron-Cousins B
passband.  Image magnitudes on the plates are usually referred to as
$b_{j}$ magnitudes.

Both the original J survey plates and the atlas copies were taken
using a strict set of criteria to minimize systematic errors between
different plate exposures and, more importantly, to ensure that the
process was highly repeatable.  For example, survey plates were only
taken in dark time, good seeing (less that 3 arcseconds) and with the
sun $>18^{\circ}$ below the horizon to avoid astronomical twilight.
During the exposure, each plate was held in a curved holder and was
flushed with nitrogen to reduce the effects of differential
desensitisation over the plate$\footnote{We note here that a few EDSGC
plates may have been taken before nitrogen--flushing was instigated}$.  The developing and copying of the
plates were also carried out with a high degree of consistency to
ensure that systematic differences between plates were not
introduced.  Finally, quality control checks were carried out on all
plates and each was graded either A, B or C.  Therefore, this atlas
represented the most homogeneous catalogue of photographic plates
available at that time.  The reader is referred to the UK Schmidt
Telescope Unit Handbook (1983) for a description of the procedures
used in taking the original plates, Cannon et al. (1978) for full
details of the processing of the plates and Bruck \& Waldron (1984) for
a discussion of the copying process.

In total, 60 grade A plates were used in the EDSGC and these are
listed in Collins, Nichol \& Lumsden (1992; CNL92).  These plates were
extracted from the ESO/SERC Atlas held in the UK Schmidt Library at
the Royal Observatory Edinburgh and scanned by COSMOS, also at the
Royal Observatory Edinburgh.  The COSMOS machine was a high-speed
flying spot micro--densitometer, specifically designed and constructed
for scanning astronomical photographic plates (details of COSMOS can
be found in MacGillivray \& Stobie (1984) while the characteristics of
its descendent, SuperCOSMOS, are to be found in Miller et al. 1991) .
The COSMOS machine originally sat on a plinth driven 35 feet into the
ground which separates it from the rest of the building thus
preventing vibrations of any sort from effecting the scans.  In
addition, COSMOS and the UKST plate library were kept in a dust-free
environment, thus reducing the chances of contamination on the plates.

Plates were fed into COSMOS with their south side at the top of the
plate carriage holder and raster scanned using a beam of light from a
cathode ray tube of width 8 microns.  As the plate moved in the y
direction, the beam scans in the x direction with a pixel size of 16
microns.  In total, an area of $287{\rm mm}\times287{\rm mm}$ was
scanned each time which corresponds to an area of
$5.35^{\circ}\times5.35^{\circ}$ on the sky.  For each pixel, a
transmission value was calculated by comparing the light measured
passing through the plate with a reference signal.  This in turn, was
converted into a measured intensity using a Baker density calibration
curve (MacGillivray \& Stobie 1984).

COSMOS was operated in two modes; mapping and threshold.  The first
recorded all the pixel information on a plate and therefore required a
vast amount of computer storage.  The second recorded only pixels that
were a certain percentage above the sky background of the plate (the
typical sky background for these plates is $\simeq22.3$ magnitude per
square arcsecond in the photographic $b_j$ system).  This is
determined beforehand by scanning the plate with a much lower
resolution (32 microns) to measure the large scale variations over the
plate which are usually due to a combination of large nearby stars,
vignetting and differential desensitisation.  After scanning, the
pixel data was passed into the COSMOS image analyzer (see Beard,
MacGillivray \& Thanisch 1990) which connects all adjacent pixels
producing a final set of objects for each plate.  Each image was then
assigned 27 individual image parameters such as the image magnitude,
position and both the intensity weighted and unweighted moments of the
pixel distribution (Stobie 1980 and Section \ref{avail} below).
Plates used in the EDSGC were scanned in threshold mode, with a
threshold of 8\% to 10\% above the sky background.  This ensured that
there were approximately the same number of objects in each scan and
was found by MacGillivray \& Dodds (1982) to be an acceptable level
for maximizing the number of true images compared to `noise' images.

The magnitudes returned by COSMOS were isophotal magnitudes and the
threshold quoted above for the EDSGC corresponded to a final isophote of
25 magnitudes per square arcsecond in  $b_j$.  For images brighter than 
$b_j=20.5$, the magnitudes were effectively total magnitudes (MacGillivray \&
Dodds 1982).  In addition, the measured magnitudes depend upon the sky
background magnitude of the scanned plate and adopting a fixed
detection threshold for all the plates introduced large variations
between the zero-point magnitudes of the plates.  Therefore, it was
imperative to obtain external photometry to calibrate the magnitude
scale of each plate.

All the COSMOS scans used in the EDSGC were analyzed using the COSMOS
de-blending software (Beard et al. 1990), which involved
re-thresholding each image in intensity space at 8 progressively
higher thresholds in search of saddle-points in the image's intensity
distribution.  If such saddle-points were found, the separate peaks
were fitted by Gaussians and split into their daughter images.  At
the SGP, the number of blended objects was found to be $\simeq10\%$ at
all magnitudes and implementing the de-blending software substantially
reduced this number {\it e.g.} at a surface density of 20 galaxies per
square arcminute ({\it e.g.} the cores of rich clusters) the number
of real objects detected increased by over 30\% because of de-blending
(Heydon-Dumbleton et al. 1989).  De-blending of the images was vital to
the EDSGC for two reasons.  First, faint star-star mergers near the
plate limit imitated galaxies because their combined shape appeared
elliptical and they had a lower surface brightness compared to a
single star.  If these were not de-blended, then there would have been
a significant contamination of false galaxies at faint magnitudes.
Secondly, the EDSGC was used in the construction of an automated
cluster catalogue (Section \ref{science}).  If de-blending had not been
implemented, the cores of many rich clusters would have appeared as single
large objects and therefore, would not have been detected by the
automated cluster detection algorithm.

\subsubsection{Star-Galaxy Classification}
\label{star}

A detailed discussion of the COSMOS star-galaxy separation criteria
can be found in Heydon--Dumbleton et al. (1989). However, for 
completeness, we given an overview of the techniques here.

A COSMOS scan of a typical Schmidt plate contains on average 200,000
objects.  To the plate limit, over 90\% of these objects are stars
which clearly must be removed to produce a reliable and meaningful
galaxy catalogue.  For the EDSGC, this was achieved using 3
parameters defined during the scanning of the plates.  Each parameter
worked over a different magnitude range and their combined effect
covered the full range of magnitudes observed on a plate.

These 3 classifiers were: The G classifier which works for magnitudes
brighter that $b_j\simeq16$ and effectively measures how well an
object fills the ellipse fitted to it (for stars brighter than
$b_j\simeq16$, diffraction spikes dominate the fitted ellipse but they
have a relatively small area, while galaxies at this magnitude tend to
fill their fitted ellipse): The Log-area classifier which works at
intermediate magnitudes ($16<b_j<19.5$) and relies on the fact that
galaxies have a lower surface brightness than stars: The S classifier
which works for magnitudes fainter than $b_j=19.5$ by comparing the
measured size of the image to that expected for a Gaussian
point--spread function.

Near the limit of the plates, star--galaxy separation became extremely
difficult.  Therefore, COSMOS scans used in the EDSGC were cut at a
COSMOS magnitude of -1.0, which roughly corresponded to $b_j\sim21$
for most plates.  To test the reliability of the star-galaxy
separation, visual checks were carried out on 5 plates spread across
the EDSGC.  For each plate, 300 classified galaxies and 300 classified
stars, over a broad range in magnitude, were randomly selected and
visually inspected.  The result of this test was a $>95$\%
completeness for the galaxies at all magnitudes with $<10$\%
stellar contamination.  Similar results were obtained for visual
checks on images selected before and after de-blending.  For a much
fuller discussion of these classifiers, their effective magnitude
ranges and the visual tests the reader is referred to either
Heydon-Dumbleton et al. (1989) or Heydon-Dumbleton (1989).

\subsubsection{Photometric Calibration}

As stated above, COSMOS only returned the magnitude of objects
relative to the background magnitude limit of the plate.  Therefore,
it was essential to obtain external photometry to calibrate all the
galaxies in the EDSGC by determining the sky background magnitude or
zero-point magnitude of the plates.  For galaxies, the relationship
between the COSMOS magnitude and the $b_j$ magnitude is linear over a
wide range of magnitudes and can be represented by $b_j=m_{cosmos} +
m_{sky}$ where $m_{cosmos}$ is the instrumental magnitude (and
negative) and $m_{sky}$ is the constant sky background of the plate
(see below and Table \ref{sequence}).

The EDSGC was calibrated using CCD direct images obtained at CTIO and
SAAO.  In total, 31 calibration sequences were taken across the whole
EDSGC and were spaced in a ``checker board'' fashion i.e. plates
either had a sequences on them or overlapped with two or more plates
with a sequence on.  Each CCD frame was centered on a loose cluster
which resulted in $\sim15$ usable galaxies per frame for the
calibration.  For the plates with a sequence, the COSMOS magnitudes of
the observed galaxies were plotted against their CCD magnitudes and
from the fit (with the slope fixed at 1.0) $m_{sky}$ was
obtained. These zero-points were then used to calibrate the whole of
the plate with a uncertainty of $\sim 0.05$ magnitudes. In Table
\ref{sequence}, we present our measured $m_{sky}$ values for the 31
EDSGC plates with a CCD sequence.  Plates marked with an asterix have
an uncertain calibration and should be viewed as preliminary at
present.  For plates without a sequence, $m_{sky}$ was calculated
using the galaxies in the overlap regions with plates with a sequence
(typically 1000-3000 galaxies in each overlap).  On average, each
uncalibrated plate overlapped with 2-3 calibrated plates which
prevented erroneous calibrations from propagating through the survey
(see our discussion of edge plates below).

The histogram of measured plate magnitude offsets between adjacent
plate zero--points is presented in CNL92 and the best fit Gaussian to
this distribution has a dispersion of 0.08 magnitudes.  This implies a
calibration uncertainty of 0.05 magnitudes on each plate which is
within the limits set by Geller, Kurtz \& de Lapparent (1984) for
plate matching errors for any new galaxy catalogue used in measuring
the large scale distribution of galaxies.  Once again, the reader is
referred to Heydon-Dumbleton (1989) and Nichol \& Collins (1993) for a
more detailed description of the photometric calibration of the EDSGC
and a fuller discussion of the errors.

\subsection{Final Catalogue}

\subsubsection{Spurious Detections and Systematic Errors}

Before the plates were joined to form one homogeneous catalogue, the
individual COSMOS scans were cleaned of spurious objects.  The main
source of such objects was from the accidental de-blending of star
halos around bright stars in the field ($b_j<12$) which tended to
mimic rich clusters of galaxies.  To combat this effect, the areas
around stars brighter than $m_{cosmos}=-9$ were removed or ``drilled''
with a radius of 6.7 arcminutes (UKST handbook 1983).  The exact
magnitude used to select the stars was obtained from the plots of the
G parameter against magnitude.  In total, 553 such ``drill holes''
were made and a full list of these holes can be found on the EDSGC
webpage.

Other spurious objects were produced by ghosts from very bright stars
($b_j\sim6$) in the field, satellite trails, large nearby galaxies and
dense star clusters.  Therefore, we plotted the galaxy distribution on
each plate and visually inspected these plots for any suspicious
features ({\it e.g.}  very dense clusters of galaxies etc.). Such
features were then inspected on the original glass plates to determine
if they were real astronomical objects or errors as discussed
above. If they were spurious, the data around these objects was
interactively removed from the catalogue.  In particular, Fields 404,
356 \& 469 had large star ghosts which were drilled with a radius of 40
arcminutes and Field 466 had a large de-focussed region in the
northwest corner of the plate.  This area was removed as no
alternative plate existed with a comparable overall quality.  In
total, 10 interactive drill holes were used in the EDSGC and their positions,
with comments, are given in Table \ref{drill}.

Another source of contamination in the catalogue came from satellite trails.
From the visual checks of the galaxy distribution of the plates, 11
plates were found to have this form of contamination.  Combined with
this, there were also spurious galaxies associated with the de-blending
of diffraction spikes of stars.  Both these types of spurious image
were removed using the fact that they had preferentially aligned
position angles i.e. diffraction spikes of stars were aligned with the
edges of the plate and therefore, their position angles were always
$0^{\circ}$, $90^{\circ}$ or $180^{\circ}$.  In addition, these images
have high ellipticities and filled their fitted ellipse extremely
well.  Therefore, on the plates in question, all detected objects were
plotted in the eccentricity (minor over major axis) versus position
angle plane which cleanly separated the satellite trails from the
majority of other objects.  A cut was made in eccentricity (typically
between 0.3 and 0.4) and all objects with a lower value than this cut
were plotted in the log--area versus magnitude plane ({\it i.e.} as
used for star--galaxy separation). As discussed above, the real galaxy
population stood away from the spurious population and the two were
separated simply using a straight line.  This method was very
effective at removing the satellite trails and diffraction spikes,
with typically 500 images removed on each plate.  As a check, the
objects both rejected and re-accepted into the catalogue by this
methodology were visually inspected on 5 of the 11 affected plates.
Of rejected images, $\sim8\%$ were galaxies, while for the re--accepted
images, over 95\% were galaxies.  Due to the success of this method in
removing residual diffraction spikes, it was performed on all the
plates in the EDSGC and on average $\sim100$ objects were removed per
plate.  Visual checks of these rejected objects were in agreement with
the numbers quoted above.

Once all the plates had been photometrically calibrated and cleaned of
spurious objects, the individual COSMOS scans were mosaiced together
to produce a final homogeneous catalogue of galaxies.  This was
achieved interactively because the extent of the overlap between
different pairs of plates varied extensively.  The western edges of
the UK Schmidt plates suffered the worst vignetting and
desensitisation (UKST handbook 1983), while the northern edges
displayed a systematic excess of faint galaxies which seemed to be a
problem with the COSMOS calculation of the plate background intensity
near that edge. This systematic excess was not due to the image
classification or the photometric calibration because the effect was
still present in the raw COSMOS scans (however, we note it only
effected the first few lanes of the COSMOS scan and was therefore,
only effected a small area of the plate).  The EDSGC plates were
attached together with a preference towards the eastern edges of the
plates into long strips of fixed declination, which were then added
together with a preference given towards the southern edges of these
strips. 

\subsubsection{Areal Coverage}
\label{areal}

The EDSGC contains 1,495,877 galaxies taken from 60 Schmidt survey
fields as listed in Table 2 of CNL92.  The largest contiguous region
within the EDSGC (avoiding the complicated plate boundaries of the
EDSGC) is given by: $3^{hrs}$ to $22^{hrs}$ (through $0^{hrs}$) and
$-23^{\circ}$ to $-42^{\circ}$ (B1950).  This gives a total area of
$1182{\rm deg^2}$ and should be used as the EDSGC window function {\it
i.e.} we recommend users confine themselves to this region for all
statistical analyses. This was the region used by CNL92 for the galaxy
angular correlation function.

\section{Availability of the EDSGC}
\label{avail}

The EDSGC is available for download over the internet from
http://www.edsgc.org. On this webpage, there are two
catalogue available: The first is a smaller version of the EDSGC with
only 7 attributes (see Table \ref{small}) and the full EDSGC with 27
attributes (see Table \ref{large}). In Tables \ref{small} and
\ref{large}, column 1 gives the name of the attribute, Column 2 gives
the units of the attribute and Column 3 is a short description of the
attribute. It is worth re-iterating here that the plate scale of the UK
Schmidt Telescope is 67.12 arcseconds per millimeter giving 1.07
arcseconds per COSMOS pixel. These numbers will allow users of the
EDSGC to convert some of the catalogue attributes into
astronomically meaningful values.

\section{Supplementary Information on the EDSGC}
\label{sup}

Over the past decade, there has been significant auxiliary information
gathered in the EDSGC area. Such data includes the Las Campanas
Redshift Survey (LCRS; Shectman et al. 1996), the ESO Nearby Abell
Cluster Survey (ENACS; Katgert et al. 1996), the ESO Slice Project
(ESP; Vettolani et al. 1997) and the Durham/UKST Redshift Survey
(DURS; Ratcliffe et al. 1998). The former two surveys provide
serendipitous information while the latter two redshift surveys use
the EDSGC, or a subset of the survey, as their input catalogues. In
total, there are 12,888 EDSGC unique galaxy redshifts from these four
surveys (5996 from LCRS, 3836 from ESP, 2501 from DURS and 907 from
ENACS) and we make these data available on the main EDSGC webpage (see
http://www.edsgc.org).  Only 176 EDSGC galaxies have more than one
redshift measurement from these four different redshift surveys and,
for these galaxies, the mean absolute difference in redshift is
$71{\rm km\,s^{-1}}$ (the largest observed difference is $306{\rm
km\,s^{-1}}$).

\section{External Checks}
\label{check}

\subsection{Photometry}

Extensive checks of the internal photometric calibration of the EDSGC
were carried out in Heydon-Dumbleton (1989). These checks centered
around the examination of galaxies in the overlap regions of plates
and on the galaxy number counts as a function of right ascension and
galactic latitude. These internal tests showed no evidence for
systematic calibration errors in the EDSGC.

In this section, we investigate external checks of the EDSGC
photometry since as demonstrated in Nichol \& Collins (1993)
systematic plate--to--plate photometric errors can significantly
affect the form of the galaxy angular correlation function (see
CNL92).  Therefore, it is important to have a
homogeneous magnitude system across the whole EDSGC.

\subsubsection{External $b_j$ CCD data}

In Nichol (1993), we presented two external checks of the EDSGC
photometry.  The first of these used the published CCD photometric
calibration sequences of the APM galaxy survey (Maddox, Efstathiou \&
Sutherland 1990a). At the bright end ($b_j<17$), the scatter in
magnitude is large with a typical difference between the two surveys
of $\pm0.2$ magnitudes.  However, the majority of the comparison data
are for galaxies fainter than a mean magnitude of $b_j=18$.  Over a
range of 3 magnitudes, the data are consistent with the expected
scatter due to the APM \& COSMOS machine measuring errors ($\sim0.1$
magnitudes).  The main result of this comparison is an overall 0.2
magnitude shift at all magnitudes between these two independent
surveys, with the APM being the fainter of the two. A large part of
this discrepancy could be the difference between total (APM) and
isophotal (EDSGC) magnitude measurements

In terms of the correlation analysis presented in CNL92, a simple
shift in the global magnitude calibration between the APM and COSMOS
catalogues is not a significant problem, since the angular correlation
function is scaled by the observed number density of galaxies in the
catalogue.  A potentially more serious problem would be the existence of
a variable magnitude shift as a function of plate position or right
ascension, as this would introduce large scale gradients.  To test for
this, the mean magnitude shift of EDSGC plates with an APM sequence
were investigated as a function of Right Ascension. As discussed in
Nichol (1993), this analysis showed no signs of a systematic variation
in the 0.2 magnitude shift between the two surveys across the whole
EDSGC survey.  This is supported by a linear fit to the data, which
gives $\Delta m_{APM-COSMOS} = (0.18\pm0.04)-(1.6\pm2.8)\times
10^{-2}\alpha$, where $\alpha$ is the right ascension of the EDSGC
plate center (24 hours was subtracted for RA coordinates greater than
12 hours).  This relationship predicts an offset of only 0.07
magnitudes between the two ends of the EDSGC, with a one sigma upper
offset limit of 0.21 magnitudes.

In addition to the APM CCD sequences, Nichol (1993) used two CCD
calibration sequences used by Matthew Colless in his study of the
cluster luminosity function (Colless 1986). This comparison however
does not show any systematic displacement between the EDSGC and
Colless zero--points and is fully consistent with the expected scatter
about zero given the COSMOS machine measuring error.  In addition, the
two Colless sequences (on Fields 349 and 405) are separated by nearly
2 hours in right ascension and show no evidence for a systematic shift
in the EDSGC magnitude zero--points.

\subsubsection{Other External Photometry Data}

As outlined in Section \ref{sup}, there are 5996 galaxies in common
between the Las Campanas Redshift Survey (LCRS; Shectman et al. 1996)
and the EDSGC.  These galaxies were matched using a search radius of 4
arcseconds -- centered on the EDSGC galaxy coordinates - which seemed
appropriate based on our separation analysis. If
an EDSGC galaxy had two, or more, possible match-ups, we forced there
to be a unique match--up by simply taking the closest LCRS galaxy.

In addition to providing redshift measurements, the LCRS also
published (Thuan-Gunn) r--band photometry on all their galaxies which
allows us to study the large--scale distribution of galaxy colors
($b_{j}-r$) across the entire EDSGC right ascension coverage {\it
i.e.} the EDSGC overlaps with the $-39^{\circ}$ and $-42^{\circ}$
declination strips of the LCRS which span from $\simeq 21^{hrs}$ to
$\simeq 4^{hrs}$ in right ascension. Therefore, this provides a
significant external check of our photometry since, on average,
galaxies should have the same $b_{j}-r$ color across the survey,
ignoring reddening effects, or real large--scale structure in the
universe. We note here that we are not making any claims about the
absolute $b_{j}-r$ color of galaxies as this would require a detailed
analysis of both the EDSGC and LCRS aperture photometry; we are simply
examining the large--scale trends in the $b_{j}-r$ color as a function
of right ascension and original plate number

\begin{figure}
\centerline{\psfig{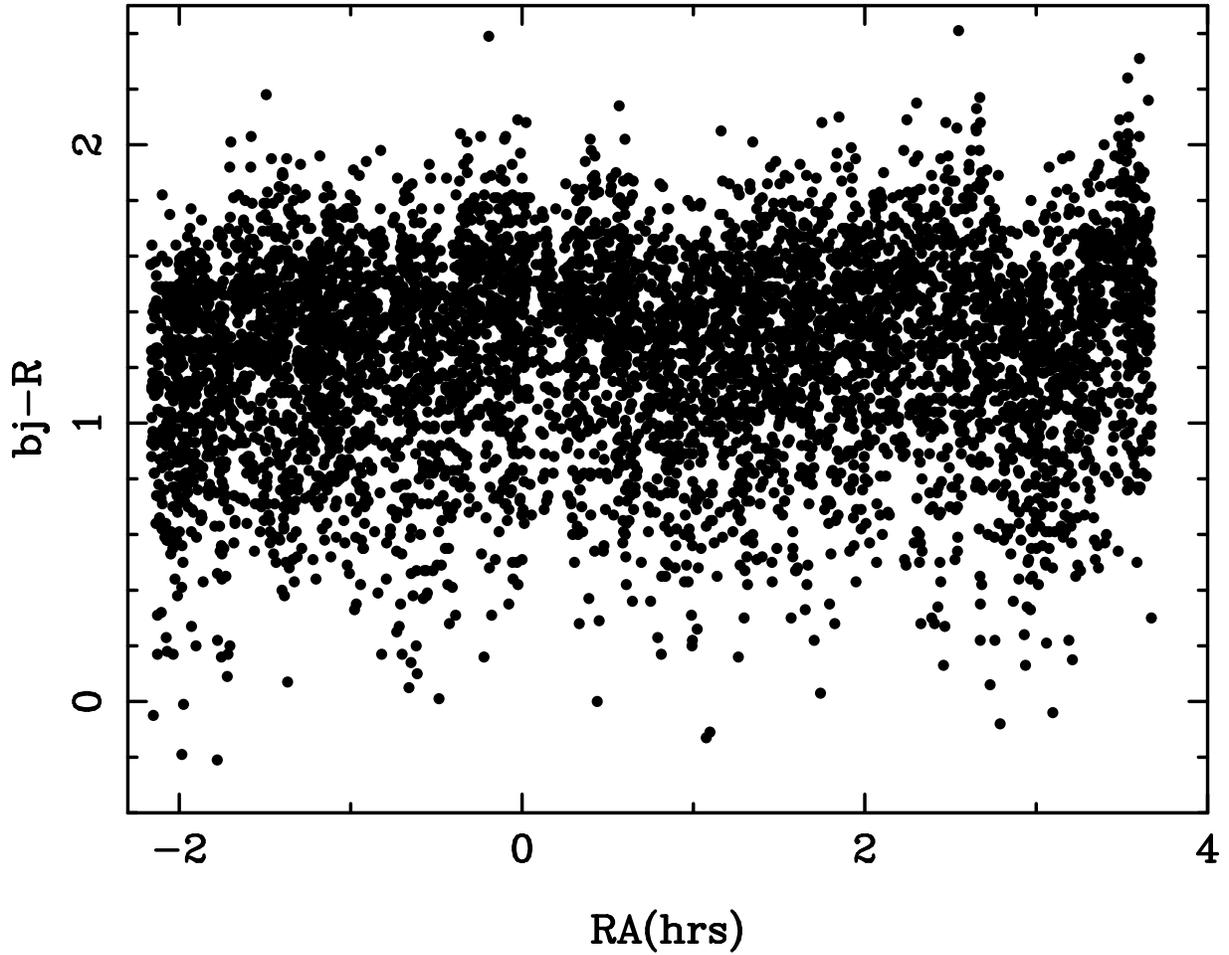}}
\caption{We present here the $b_{j}-r$ color for all 5996 galaxies in
common between the EDSGC and LCRS as a function of the EDSGC right
ascension of the galaxy.  We have made no cuts on the data.
\label{fig1}}
\end{figure}

In Figure \ref{fig1}, we plot the $b_{j}-r$ color of all galaxies as a
function of right ascension. This plot immediately demonstrates that
there is no large--scale systematic error between these two
independent photometric systems {\it i.e.}  the mean galaxy $b_{j}-r$
color is a constant over 70 degrees in right ascension for this joint
survey.  This is more clearly shown in Figure 2a which plots the mean
and standard deviation of the $b_{j}-r$ color versus right ascension
for the EDSGC--LCRS sample. The scatter between these mean points is
significantly smaller than the intrinsic scatter seen in the
individual galaxy colors (Figure 1) and is fully consistent with the
sum of errors on the individual photometric measurements {\it i.e.} as
discussed above and in CNL92, the rms plate--to--plate scatter for the
EDSGC is $\Delta m\simeq 0.08$ (which does not include the
photographic measurement error) while the standard deviation of
pairwise galaxy magnitude differences for LCRS galaxies which were
measured twice on overlapping ``bricks''$\footnote{ A brick is a
$1.5^{\circ}$ x $3^{\circ}$ region as discussed in Shectman et
al. 1996}$ is $\sigma_m = 0.10$, for $16.0 < r \le 17.0$, and
$\sigma_m = 0.17$, for $17.0 < r \le 18.0$ (see also Figure 3 of
Shectman et al. 1996).

To investigate this further, we have plotted in Figure 2b the mean
$b_{j}-r$ color for the EDSGC--LCRS galaxies as a function of their
original EDSGC plate identification and right ascension (there are only
LCRS data on 14 of the original 60 EDSGC plates).  In this plot, we
present the mean EDSGC--LCRS galaxy color for each plate as a function
of the mean right ascension of that plate. We have broken the data
into the two different LCRS declination slices, {\it i.e.}
$-39^{\circ}$ and $-42^{\circ}$, since the former of these two runs
through the center of the EDSGC plates (where the calibration is
expected to be best), where the latter runs across the southern edge
of the EDSGC (as discussed in Section \ref{areal}, we cut the EDSGC at
$-42^{\circ}$ for statistical analyses).

Figure 2b once again demonstrates that there is no large--scale
gradient between the two photometric systems. The amplitude of the
scatter seen between the mean EDSGC--LCRS galaxy color for
$-39^{\circ}$ declination strip is fully consistent with the rms
scatter quoted above and in CNL92.

We note here that the larger scatter seen in Figure 2b for the
$-42^{\circ}$ LCRS declination strip is probably due to photographic
plate edge effects. Such edge effects have little consequence on the
science of the EDSGC, as outlined in Section \ref{science}, since in
most cases the data was cut to avoid such edge effects as discussed in
Section \ref{areal}. The same is true for plates 300, 301 \& 344
(regardless of LCRS declination slice) since these three plates are at
the corners of the EDSGC where the number of plate overlaps is
significantly smaller than for other plates; plate 300 only overlaps
with plate 301 which in turn only overlaps with plates 357 \& 299. For
plate 301, the measured magnitude offset with these overlapping plates
is $\Delta m_{301-357}=0.35$ and $\Delta m_{301-299}=0.11$ {\it i.e.}
the average magnitude difference for galaxies in common between two
plates. This may explain the larger scatter seen for plates 300 \& 301
since they are somewhat isolated from the rest of the EDSGC (they are
on the southeast corner of the EDSGC). Similarly, Plate 344 is on the
southwestern corner of the EDSGC and also suffers from a smaller
number of plate overlaps with other plates; it only overlaps with
plate 405 ($\Delta m_{344-405}=0.028$) and plate 345 ($\Delta
m_{344-345}=-0.05$).

We note that the feature in Figures 1 \& 2 near Plate 293 could be due
to real large--scale structure. As shown in Guzzo et al. (1992), this
area of the EDSGC coincides with both the Sculphor and BS1
superclusters; two of the largest superclusters presently known.

Finally, we have also carried out a similar analysis using the 907
galaxies in common between the ESO Nearby Abell Cluster Survey (ENACS;
Katgert et al. 1996) and the EDSGC (ENACS provides a R--band magnitude
for all their galaxies).  The results are the same as above with the
LCRS in that the mean $b_{j}-R$ color for EDSGC-ENACS galaxies is a
near constant as a function of right ascension.

\subsection{Star--Galaxy Separation}

As discussed in Section \ref{sup}, the EDSGC has been used for several
large galaxy redshift surveys. A by--product of this work is an
external check on the star--galaxy separation techniques of the EDSGC
since stars are easy to spectroscopically differentiate from galaxies.
The best test of this comes from the ESO Slice Project (Vettolani et
al. 1997) since they targeted for spectroscopic observations all
$b_j\le19.4$ EDSGC galaxies between a right ascension of $22.5^{h}$
and $1^{h}\,20^{m}$ at a mean declination of $-40.25^{\circ}$.  From
this sample, which is 90\% complete in observations, only 12\% of
EDSGC galaxies were stars. This is slightly higher than we observed
from our visual checks (Heydon-Dumbleton et al. 1989 quotes $<10\%$)
of the star--galaxy separations discussed above in Section \ref{star}.
 
\section{Discussion}
\label{science}

In this section we give a brief review of the science to originate
from the EDSGC:

\subsection{Galaxies}

The angular correlation function ($w(\theta)$) of the EDSGC and its
stability to Galactic extinction and photometric calibration errors
are given in CNL92 and Nichol \& Collins (1993) respectively.  These
results confirm the high amplitude of the angular correlation function
found by the APM survey (Maddox {\it et al.} 1990b) and help
demonstrate that the previous results pertaining to the Lick galaxy
catalogue (Groth \& Peebles 1977) were in error. Higher order
correlations have been analysed by Szapudi, Meiksin \& Nichol (1996)
which are in good agreement with predictions from N-body simulations
and suggest that the galaxies are reliable tracers of the underlying
mass distribution. A comparison of the large-scale clustering in the
APM and EDSGC galaxy surveys was performed by Szapudi \& Gaztanaga
(1998). On large angular scales the main difference between counts in
cells is attributable to the smaller area, and therefore greater edge
effects, of the EDSGC compared to the APM. On small scales there is
evidence that the EDSGC follow N-body simulations more closely than
APM which could be a result of the EDSGC de--blending algorithm.
However, none of the differences detected effect the results on
large-scale structure or their interpretations as published by the two
groups.

The EDSGC has been the parent galaxy catalogue for two extensive
redshift surveys: the ESO Slice Project (ESP) and the Durham/UKST
Galaxy Redshift Survey. The ESP is an $85\%$ complete galaxy redshift
survey consisting of $\sim 3300$ galaxies brighter than $b_j=19.4$ in
a 23.3 square degree region close to the South Galactic Pole
(Vettolani {\it et al.} 1997).  Most notable are the results on the
luminosity function and mean galaxy density (Zucca {\it et al.} 1997)
which provide clear evidence for a ``local'' underdensity in the
galaxy distribution out to $\sim 140$ h$^{-1}$ Mpc. Along with the Las
Campanas (Shectman {\it et al.} 1996) and Stromlo-APM (Loveday et
al. 1996) redshift surveys, the ESP project provides one of the most
accurate estimates of the faint end of the galaxy luminosity
function. The Durham/UKST Galaxy Redshift Survey consists of a
one-in-three sampled survey of $\sim2500$ galaxies to $b\leq17$
covering the EDSGC area (Ratcliffe {\it et al.} 1998). The results of
this survey suggest that both the structures on scales 50-100
h$^{-1}$Mpc and the smaller scale pairwise galaxy velocity dispersion
show significant power in excess of the predictions of the standard
cold dark matter model.  Finally, Knox, Huterer \& Nichol (2000) have
used this extensive redshift data to compute an angular power
spectrum, and its covariance matrix, for the whole EDSGC data and
compare these results to cosmological predictions.

\subsection{Clusters}

The work on galaxy clusters from the EDSGC has been particularly
productive and includes: publication of the first machine-based
cluster catalogue -- the EDCC (Lumsden {\it at al} 1992); results on
the spatial distribution and correlation function of clusters (Guzzo
{\it et al.}  1992, Nichol {\it et al.} 1992); an analysis of the
alignments of galaxy clusters (Martin {\it et al.} 1995); a study of
the cluster galaxy luminosity function from the largest single
database ever used to study this problem (Lumsden {\it et al.} 1997);
a comprehensive redshift survey of rich clusters (Collins {\it et al.}
1995). In addition, the EDSGC has been used recently to study the
local space density of optical clusters (Bramel, Nichol \& Pope 2000)
and the relationship between optical and X--ray luminosities of
clusters (Miller, Melott \& Nichol 2000).

\section{Acknowledgments}

We are grateful to the COSMOS and UK Schmidt units at the Royal
Observatory Edinburgh and Anglo--Australian Observatory for their
continued support of the EDSGC. We would also like to thank PPARC for
their financial support of the EDSGC. We are indebted to Neil
Heydon-Dumbleton and Luigi Guzzo for their help and advice over the
lifetime of this project and we thank Matthew Colless for making his
CCD calibration data available to us. Finally, we thank Marion Schmidt
of the NASA/IPAC Extragalactic Database (NED) for his continuing
checks of the EDSGC. This work was sponsored in part by NASA grant
NAG5-3202 (RCN).

\newpage


\begin{figure}
\centerline{\psfig{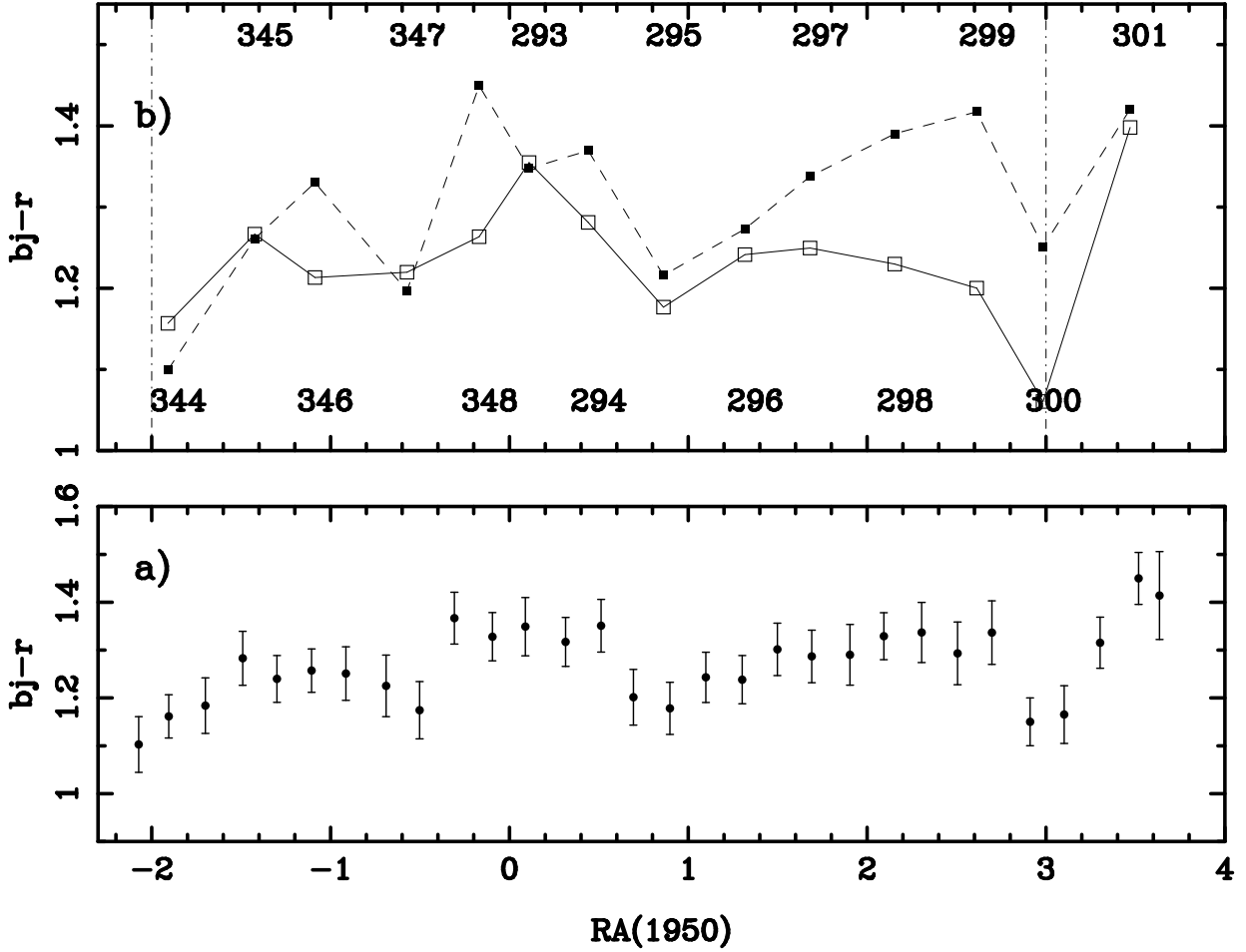}}
\caption{In panel a), we present the mean, and standard error on the mean, of
the $b_{j}-r$ color for all EDSGC--LCRS galaxies in the EDSGC
magnitude range $17<b_j<20$ as a function of the EDSGC right ascension
of the galaxy. In this magnitude range, the EDSGC is considered to be
free of systematic problems like saturation of galaxy cores (see
Lumsden et al. 1997) as well as being well above the flux limit of the
photographic plates. In panel b), we plot the same data as shown in
panel a) but now as a function of plate identification number {\it
i.e.}  we present the mean $b_{j}-r$ color for all EDSGC--LCRS
galaxies on a given plate. We plot the color data versus the mean
right ascension of the plate. The solid square symbols are for
galaxies in the $-42^{\circ}$ LCRS declination slice while the open
square symbols are for the $-39^{\circ}$ LCRS declination slice. The
plate identification number for each pair of points is given either
above or below the data (to avoid overcrowding on the plot). The lines
are presented to aid the reader and have no scientific rational.
The dot-dashed vertical lines are the limits of the EDSGC as discussed
in Section \ref{avail}.
\label{fig2}}
\end{figure}

\clearpage

\begin{table}
\caption{The EDSGC plates with CCD Sequences 
\label{sequence}
}
\begin{tabular}{cc|cc|cc}\hline\hline
Plate No. & $m_{sky}$ & Plate No. & $m_{sky}$ & Plate No. & $m_{sky}$ \\ \hline
293&$22.14\pm0.04$&  294&$22.31\pm0.04$&   296&$22.20\pm0.04$ \\
298&$22.41\pm0.05$&  300&$22.34\pm0.05$&   342&$22.35\pm0.03$ \\
344&$21.77\pm0.06$&  346&$22.17\pm0.03$&   347&$22.24\pm0.05$ \\
348&$22.25\pm0.06$&  349&$21.77\pm0.02$&   351&$22.48\pm0.06$ \\
353&$22.39\pm0.04$&  355&$22.00\pm0.07$&   405&$22.22\pm0.04$ \\
407&$22.14\pm0.04$&  410&$22.58\pm0.03$&   411&$21.61\pm0.04$ \\
412$^{\star}$&$22.40\pm0.07$&414&$22.27\pm0.05$&416$^{\star}$&$22.03\pm0.03$ \\
466&$22.10\pm0.05$&  468&$22.43\pm0.06$&   469&$22.00\pm0.05$ \\
470&$22.13\pm0.05$&  471$^{\star}$&$22.42\pm0.05$&  472&$22.40\pm0.04$ \\
474&$21.97\pm0.04$&  477&$22.11\pm0.04$&   531&$21.93\pm0.04$ \\
533&$21.95\pm0.04$&     &              &      &               \\ \hline\hline
\end{tabular}
\end{table}

\begin{table}
\caption{Ten Interactive Drill Holes within the Boundaries of the EDSGC
\label{drill}
}
\begin{tabular}{llc|l}\hline
RA (hrs)& DEC(Degrees) & Drill Diameter (arc mins) & Comments \\ \hline\hline 
22.9142 & -29.89392 & 40.90 & HD 216956, a bright star\\
23.0850 & -30.04382 & 40.90 & large plate error\\ 
23.5302 & -36.37440 & 5.6   & IC5332\\ 
23.9943 & -30.00320 & 8.61  & HD 224990, a bright star\\ 
0.08746 & -37.92240 & 18.00 & NGC 0300, a bright galaxy\\
0.09624 & -34.00000 & 25.00 & the Sculpter Dwarf Elliptical\\ 
0.8373  & -31.46767 & 6.48  & NGC289/ARP 81\\ 
0.8390  & -26.85695 & 12.90 & Near NGC 288/Globular cluster\\ 
0.8824  & -37.82746 & 9.75  & Near NGC 0300\\
2.6292  & -34.71350 & 40.00 & the Fornax Dwarf Spheroidal\\
\hline\hline
\end{tabular}
\end{table}

\begin{table}
\caption{The Seven Attributes in the Small Version of the EDSGC
\label{small}
}
\begin{tabular}{ll|l}\hline
Name      & Units         & Description \\ \hline\hline
RA        & Hours         & Right Ascension (B1950)\\     
DEC       & Degrees       & Declination (B1950)      \\
MAGNITUDE & Magnitudes    & Calibrated $b_{j}$ magnitude\\
PLATE ID  & No units      & The UK Schmidt Field ID number\\
IMAJAX    & Arcseconds    & Intensity weighted semi--major axis\\
IMINAX    & Arcseconds    & Intensity weighted semi--minor axis  \\
POSANGLE  & Degrees       & Position angle on the sky\\ \hline\hline
\end{tabular}
\end{table}

\clearpage

\begin{table}
\caption{The Twenty Seven Attributes in the Full EDSGC
\label{large}
}
\begin{tabular}{ll|l}\hline
Name      & Units         & Description \\ \hline\hline
RA        & Hours         & Right Ascension (B1950)   \\  
DEC       & Degrees       & Declination (B1950)      \\
XMIN      & 0.1 microns   & x--coordinate minimum of source\\                  
XMAX      & 0.1 microns   & x--coordinate maximum of source  \\                
YMIN      & 0.1 microns   & y--coordinate minimum of source    \\    
YMAX      & 0.1 microns   & y--coordinate maximum of source\\
AREA      & pixels        & Measured area of the source      \\          
IMAX      & intensity     & Maximum intensity above sky value  \\
COSMAGCAL & magnitudes    & COSMOS or instrumental magnitude   \\
ISKY      & intensity     & Sky intensity at centroid        \\
IXCEN     & 0.1 microns   & Intensity weighted x--centroid     \\   
IYCEN     & 0.1 microns   & Intensity weighted y--centroid\\
UMAJAX    & 0.1 microns   & Unweighted semi--major axis\\
UMINAX    & 0.1 microns   & Unweighted semi--minor axis\\
UTHETA    & Degrees       & Unweighted orientation\\
IMAJAX    & 0.1 microns   & Intensity weighted semi--major axis\\
IMINAX    & 0.1 microns   & Intensity weighted semi--minor axis  \\
POSANGLE  & Degrees       & Position angle on the sky\\
CORMAGCAL & Magnitudes    & COSMOS classification flag (internal use only)\\
SIGMA     & $\sqrt{{\rm pixels}}$ & Gaussian fit parameter (see Section 2)  \\   
IDSEQ     & No units      & Sequential ID number    \\
LOGAREA   & Log(pixels)   & Logarithm of Area (see Section 2)\\     
GEOM      & No units      & Geometric parameter (see Section 2)\\
GEOMLOG   & No units      & GEOM*LOGAREA      \\
PLATE ID  & No units      & The UK Schmidt Field ID number\\
MAGNITUDE & magnitudes    & Calibrated $b_{j}$ magnitude\\
SPARE     & No units      & Spare Attribute\\ \hline\hline
\end{tabular}
\end{table}


\begin{thebibliography}{}
\bibitem[A58]{a58} Abell, G. O., 1958, ApJS, 3, 211
\bibitem[A98]{a89} Abell, G. O., Corwin, H. G. Jr., Olowin, R. P., 1989, ApJS, 70, 1
\bibitem[B90]{beard} Beard, S. M., MacGillivray, H. T., Thanisch, P. F., 1990, \mnras, 247, 311
\bibitem[Bramel99]{Bramel99} Bramel, D. G., Nichol, R. C., Pope, A. C., 2000, \apj, accepted (astro-ph/9912275)

\bibitem[bruck]{bruck} Bruck, M. T., Waldron, J. D., 1984, Astronomical Photography

\bibitem[cannon]{cannon} Cannon, R. D., Hawarden, T. G., Sim, E., Tritton, S. B., 1978, Royal Observatory, Edinburgh (Scotland).

\bibitem[collin]{collins} Collins, C. A., Heydon-Dumbleton, N. H., Macgillivray, H. T., 1989, \mnras, 237, 7

\bibitem[collns]{solins} Collins, C. A., Nichol, R. C., Lumsden, S. L., 1992, \mnras, 254, 295 (CNL92)

\bibitem[collins1]{collins1} Collins, C. A., Guzzo, L., Nichol, R. C., Lumsden, S. L., 1995, MNRAS, 274, 1071

\bibitem[dposs]{dposs} Djorgovski, S. G., Gal, R. R., Odewahn,
S. C., de Carvalho, R. R., Brunner, R., Longo, G., Scaramella, R.,
1998, in Wide Field Surveys in Cosmology, eds. S. Colombi and
Y. Mellier

\bibitem[geller]{gelee} Geller, M. J., Kurtz, M. J., de Lapparent, V., 1984, \apj, 287, L55

\bibitem[guzzo]{guzzo} Guzzo, L., Collins, C. A., Nichol, R. C., Lumsden, S. L., 1992, ApJ, 393L, 5

\bibitem[h1]{h1} Heydon-Dumbleton, N. H., Collins, C. A., Macgillivray, H. T., 1989, MNRAS, 238, 379

 
\bibitem[h2]{h2} Heydon-Dumbleton, N. H., 1989, PhD. Thesis, Univ. of Edinburgh

\bibitem[Kat]{kat} Katgert, P., Mazure, A., Perea, J., den Hartog, R., Moles, M., Le Fevre, O., Dubath, P., Focardi, P., Rhee, G., Jones, B., Escalera, E., Biviano, A., Gerbal, D., Giuricin, G., 1996, \aap, 310, 8

\bibitem[knox]{knox} Knox, L., Huterer, D., Nichol, R.C., 2000, \apj, in preparation 

\bibitem[love]{love} Loveday, J., Peterson, B. A., Maddox, S. J., Efstathiou, G., 1996, \apjs, 107, 201

\bibitem[lums]{lums} Lumsden, S. L., Nichol, R. C., Collins, C. A., Guzzo, L., 1992, MNRAS, 258, 1

\bibitem[lums1]{lums1} Lumsden, S. L., Collins, C. A., Nichol, R. C., Eke, V. R., Guzzo, L., 1997, MNRAS, 290, 119

\bibitem[Mc]{mc} MacGillivray, H. T., Stobie, R. S., 1984, Vistas in Astronomy, 27, 433

\bibitem[mc1]{mc1} MacGillivray, H. T., Dodd, R. J., 1982, Astrophysics and Space Science, 83, 127

\bibitem[dox]{dox} Maddox, S. J., Efstathiou, G., Sutherland, W. J., Loveday, J., 1990b, \mnras, 242, 43

\bibitem[dox1]{dox1} Maddox, S. J.; Efstathiou, G.; Sutherland, W. J., 1990a, \mnras, 246, 433

\bibitem[marti]{marti} Martin, D. R., Nichol, R. C., Collins, C. A., Lumsden, S. A., Guzzo, L., 1995, \mnras, 274, 623

\bibitem[miller]{miller} Miller, C. J., Melott, A. L., Nichol, R. C., 2000, \apj, submitted (astro-ph/9912362)

\bibitem[nic]{nic} Nichol, R. C., Collins, C. A., Guzzo, L., Lumsden, S. L., 1992, \mnras, 255, 21P

\bibitem[nic1]{nic1} Nichol, R. C., 1993, PhD. Thesis, Univ. of Edinburgh

\bibitem[nic2]{nic2} Nichol, R. C., Collins, C. A., 1993, \mnras, 265, 867

\bibitem[miller1]{miller1} Miller, L., Cormack, W., Paterson, M.,
Beard, S., Lawrence, L., 1991.  "Digitised Optical Sky Surveys",
eds. H.T. MacGillivray and E.B. Thomson, Kluwer Academic Publishers

\bibitem[p]{p} Phillipps, S., Parker, Q.A., Schwartzenberg, J.M., Jones, J. B., 1998, \apj, accepted.

\bibitem[rat]{rat} Ratcliffe, A., Shanks, T., Parker,
 Q. A., Broadbent, A., Watson, F. G., Oates, A. P., Collins, C. A.,
 Fong, R., 1998, \mnras, 300, 417

\bibitem[shect]{shect} Shectman, S. A., Landy, S. D., Oemler,
A., Tucker, D. L., Lin, H., Kirshner, R. P., Schechter, P. L. 1996,
\apj, 470, 172

\bibitem[stib]{stib} Stobie, R. S., 1980, British Interplanetary Society, Journal (Image Processing), 33, 323

\bibitem[sza]{sza} Szapudi, I., Meiksin, A., Nichol, R.C., 1996, \apj, 473, 15

\bibitem[sza1]{sza1} Szapudi, I, Gaztanaga, E., 1998, \mnras, 300, 493

\bibitem[ve]{ve} Vettolani, G., Zucca, E., Zamorani, G.,
 Cappi, A., Merighi, R., Mignoli, M., Stirpe, G. M., MacGillivray, H.,
 Collins, C., Balkowski, C., Cayatte, V., Maurogordato, S., Proust,
 D., Chincarini, G., Guzzo, L., Maccagni, D., Scaramella, R.,
 Blanchard, A., Ramella, M., 1997, \aap, 325, 954

\bibitem[york]{york} York, D. G., {\it et al.} (SDSS Collaboration), \aj, submitted (technical description of the SDSS)
\end{thebibliography}
\end{document}